\documentclass[manuscript]{aastex61}

\begin{document}

\title{Classifying X-ray Binaries: A Probabilistic Approach}
\author{Giri Gopalan}
\affil{Present affiliation: University of Iceland}
\author{Saeqa Dil Vrtilek}
\affil{Present affiliation: Harvard-Smithsonian Center for Astrophysics}
\author{Luke Bornn}
\affil{Present affiliation: Simon Fraser University}

\begin{abstract}
In X-ray binary star systems consisting of a compact object that accretes material from an orbiting secondary star, there is no straightforward means to decide if the compact object is a black hole or a neutron star. To assist this classification, we develop a Bayesian statistical model that makes use of the fact that X-ray binary systems appear to cluster based on their compact object type when viewed from a 3-dimensional coordinate system derived from X-ray spectral data. The first coordinate of this data is the ratio of counts in mid to low energy band (color 1), the second coordinate is the ratio of counts in high to low energy band (color 2), and the third coordinate is the sum of counts in all three bands. We use this model to estimate the probabilities that an X-ray binary system contains a black hole, non-pulsing neutron star, or pulsing neutron star. In particular, we utilize a latent variable model in which the latent variables follow a Gaussian process prior distribution, and hence we are able to induce the spatial correlation we believe exists between systems of the same type. The utility of this approach is evidenced by the accurate prediction of system types using Rossi X-ray Timing Explorer All Sky Monitor data, but it is not flawless. In particular, non-pulsing neutron systems containing ``bursters" that are close to the boundary demarcating systems containing black holes tend to be classified as black hole systems. As a byproduct of our analyses, we provide the astronomer with public \texttt{R} code that can be used to predict the compact object type of X-ray binaries given training data.
\end{abstract}

\keywords{methods: statistical; methods: data analysis; pulsars: general; stars: black holes; stars: neutron; X-rays: binaries}

\section{Introduction}

As our ability to acquire and archive data in all fields rapidly grows, the tools for searching these data for pattern, order, and meaning need to grow commensurately.  A critical issue in this ongoing paradigm shift is that of multivariate data with complex, hidden geometric structure. Color-color or CC diagrams (which provide spectral information over different energy ranges) and color-intensity or CI diagrams (which show brightness variations for a given color) are common and easily obtained measurements that have long been used to classify X-ray binary types. \citet{wm84} plotted all X-ray binaries observed by the HEAO-1 satellite on one CC plot; they found that systems containing black holes clustered in one corner of their diagram and pulsars clustered in an opposing corner. While they found significant overlap of several classes of object in the center, they were able to use this clustering to identify new BHC candidates. In Vrtilek \& Boroson (2013; hereafter VB13), we show that when CC and CI are combined into a three dimensional CCI plot, different types of X-ray binaries (XRBs) separate into complex but geometrically distinct volumes. VB13 model the volumes crudely by computing a centroid and constructing an ellipsoid around the centroid that contains 50 percent of all points while minimizing the volume of the ellipsoid. We suggest that these diagrams provide an easily used, model-independent way to separate classes of systems, particularly systems containing black holes from those containing neutron stars or systems that can produce jets from those that cannot.  As a next step towards understanding the physical mechanisms behind this separation of compact object types, we have developed a probabilistic (Bayesian) model that provides a supervised learning approach: unknown classifications of XRBs are predicted given known classifications.  We provide the astronomer with \texttt{R} code that takes as input CCI data and outputs the estimated probabilities that a system is a black hole, pulsar, or non-pulsing X-ray binary system, in addition to standard errors for these estimates. This software provides the astronomer with more information than an off the shelf machine learning solution because such a method typically produces point estimates for the classes of the observations, as opposed to an entire distribution for the classes of the observations.

In Section 2 we describe the data used; in Section 3 we specify the models we have used for estimating the probabilities that the compact object type of an X-ray binary system is a black hole, non-pulsing neutron star, or pulsar. In Section 4 we present our results and their implications, and in Section 5 we conclude with a summary and future directions for this work.

\section{Data}

Data on X-ray binaries were obtained by the All Sky Monitor (ASM) \citep{lev96} on board NASA's Rossi X-ray Timing Explorer (RXTE), which operated continuously for nearly fifteen years. Due to as yet uncalibrated gain changes in the instrument over the last two years of its life,  we use only data obtained within the first 13 years. The MIT ASM team provides the data in three energy bins (1.3-3.0keV; 3.0-5.0keV; 5.0-12.0keV) sampled 4-8 times a day. We take one day averages and define our colors as ratios of mid to low energy range (C1) and high to low energy range (C2). The sum of the three energy bands is used to represent the intensity of the source; this value is normalized by dividing the total counts by the average of the top 1 percent of the data for any given source. These form a three-tuple consisting of the features, or background covariates in statistical terms, for each of the observations. Note that since these features are defined in terms of ratios of counts, they are unitless.  We also restrict ourselves to detections that have a signal to noise of at least five, where signal to noise is defined as the ratio of the number of counts in a particular bin to the error on the number of counts. Fig. 1 shows an example of a CCI diagram constructed with three types of X-ray binaries. 

The classification of X-ray binaries is not simple and different authors tend to use different criteria. For our training set, we used 24 systems whose classifications are consistent over numerous authors. In particular, we first considered the classifications from the catalogs of \citet{liu01,liu06}. We then used \citet{rem06} identifications of confirmed black hole systems, \citet{hom10} identifications of Z and Atoll sources, and \citet{bild97} identifications of accreting pulsars. We excluded any of the above types that were also identified as bursters as these vary from author to author. This left us with 9 systems containing confirmed black holes (Cyg X-1, LMC X-1 , J1118+480, J1550-564, J1650-500, J1655-40, GX 339-4, J1859+226, GRS 1915+105), 9 confirmed pulsars (J0352+309, J1901+03, J1947+300, J2030+375, J1538-522, Cen X-3,  Her X-1, SMC X-1, Vela X-1), and 6 non-pulsing neutron star systems (Sco X-1, Cyg X-2, GX 17+2, GX 349+2, GX 9+1, GX 9+9).

We test our model by predicting the compact object type of three groups of systems. The first contains 6 systems whose type is unambiguously classified: one confirmed black hole (LMC X-3); one non-pulsing neutron star system (GX 5-1); and 4 pulsing neutron star systems (1744-28, 0656-072, 0535+262, 0115+634). The second group of systems contain stars that are classified as both burster and atoll sources (Ser X-1, Aql X-1, 1916-053, 1608-522, 1254-69, and 0614+091) and hence have possibly ambiguous classifications. The third group of systems are sources that are either unclassified or have multiple classifications across various authors (1900-245, GX 3+1,1701-462, 1636-53, and 1700-37).

The training data set consists of 40857 observations after the preprocessing steps delineated above,  of which 13098 come from black hole systems, 25366 come from non-pulsing neutron star systems, and 2393 come from pulsing star systems. The imbalance in observations between different compact star types is due to the fact that brighter sources are more likely to be detected than weak sources, yet the brightness of sources is not uniform amongst star systems. See  Figure 2 comparing brightness for various systems.

Because observations that are less bright are inherently more likely to be below the signal to noise threshold that we utilized in preprocessing,  the data are not missing at random \citep{lit02}. The imbalance in observation types by system is problematic because visual inspection of systems of the same compact object type indicates substantial variability between systems; hence, it is prudent to ensure that the true variation of CCI values for each compact object is accurately reflected in the training set. For instance, see Figure 3 for a visualization of the variance between systems that are black holes. Additionally, the model we employ for generating multiple imputations, discussed in the next section, involves a Gaussian process and hence can be quite computationally expensive to work with, generally scaling with computations that take $O(N^3)$  time where $N$ is the number of data points in the data set. One solution to mitigate both of these issues is to subsample the training set where the probability that a particular observation is selected for inclusion in the smaller training set is inversely proportional to the total number of observations of its system in the entire training set. We sample 10 percent of the training data in this manner, without replacement, to achieve a final training set consisting of  4085 observations, of which 1486 come from black hole systems, 1465 come from non-pulsing neutron star systems, and 1134 come from pulsing neutron star systems. The  histograms in Figure 4 show the balance between various systems before and after subsampling.

\section{Models and Algorithm}
Our approach estimates the probabilities that the compact object type of an XRB is a black hole, non-pulsing neutron star, or pulsar from posterior predictions of the compact object type associated with CCI observations within the system. This approach is similar to the multiple imputations methodology developed in the context of survey analysis, which is discussed in further detail by \citet{rub96}. In summary, the multiple draws from the predictive distribution of compact object type allow the astronomer to make a judgement about her or his belief about what the compact object type of a given system is, and provide more information than a point prediction. The salient property of this approach is that it takes into account the inherent uncertainty in the prediction for each individual observation,  and a concrete illustration of the output of this methodology is displayed in the Results section. For more on  applied Bayesian data analysis, see \citet{gel13}.

The astronomer's probability model for the compact object of observations from a particular system is a trinomial model, where the relevant estimands are the probabilities that the compact object type is a black hole, non-pulsing neutron star, or pulsar. The astronomer's objective is to estimate these probabilities given the compact object type for all observations within the system. Note that by employing a model which is not a constant, the astronomer is implicitly modeling ``imputer noise": in principle the compact object type of a system should be constant for all observations from that system, and so the probability model employed by the astronomer is not physically correct. Instead it is a pragmatic solution to allow for the (inevitable) mistakes made by an imputer. Indeed, it would be unrealistic to assume any probabilistic model to be accurate all of the time.

Next, we describe the probabilistic (Bayesian) model used to generate predictions for the compact object type of CCI observations. We denote the training set as the 2-tuple $(X_{train}, Y_{train})$ where $X_{train}$ is an $N_{train}$ by 3  matrix consisting of the three CCI values of the training points, and $Y_{train}$ is a length $N_{train}$ vector of labels 1,2, or 3 corresponding to the compact object type of the system each individual data point comes from: more precisely 1 represents a black hole system, 2 represents a non-pulsing neutron star system, and 3 represents a pulsing system. The test set is denoted by $X_{pred}$ which is an $N_{pred}$ by 3 matrix that contains the CCI values of observations from a single system we would like to predict the compact object type of, and we write $Y_{pred}$ to indicate the labels for the observations from the system. The model aims to predict the unknown vector $Y_{pred}$ and from this estimate the probabilities that the compact object type is a black hole, pulsar, or non-pulsar. From this point of view, $Y_{pred}$ is the inferential object of interest, and the remaining parameters discussed are nuisance parameters, meaning that they exist for the mathematics of the probabilistic model employed but are not of ultimate inferential interest.

We introduce three independent latent variables for each compact object type (black hole, non-pulsar, and pulsar). Each latent variable has a marginal distribution that is a Gaussian process with mean 0 and covariance matrix $\Sigma$ that has a squared exponential kernel, a standard choice in the computer experiments and machine learning literature, which is discussed in detail by \citet{ras05}, i.e.
\begin{eqnarray}
\Sigma_{ij} &=& \sigma^2\exp(-||X_{i,.}-X_{j,.}||_2^2/\phi)
\end{eqnarray}
where $X_{i,.}$ denotes the $i_{th}$ row of the data matrix $X$ for which the first $N_{train}$ rows are the rows of $X_{train}$ and the rows from $N_{train}+1$ to $N_{train}+N_{pred}$ contain the rows of $X_{pred}$. Each latent variable is tied to the compact object type through a multinomial logistic link function: the probability of a particular observation $k$ to be of a class $l$ is proportional to $exp[\alpha_l + \beta_lZ_{kl}]$, where $\alpha_l$ and $\beta_l$ have independent unit normal marginal distributions, and $Z_{.,l}$ is the latent variable drawn corresponding to type $l$. Note that $Z_{.,l} \in \mathbb{R}^{N_{train}+N_{pred}}$, where the first $N_{train}$ elements correspond to the training points, $Z_t$ for short, and remaining elements correspond to the prediction test points, $Z_p$ for short. The parameter $\alpha_l$ is the mean offset for the latent Gaussian process corresponding to compact object type $l$, and the parameter $\beta_l$ is indicative of the marginal effect that each latent variable has on the propensity of a data point to be of type $l$. It is important to recapitulate that since we are interested in predicting the vector of compact object types $Y_{pred}$, these additional parameters introduced ($\alpha_l$ and $\beta_l$) in our model are ultimately marginalized out (i.e., forgotten) as nuisance parameters.
The likelihood of the observed labels given the latent variables and remaining nuisance parameters takes a multinomial form:
\begin{eqnarray}
L(\alpha,\beta, Z_t; Y_{train}) &=& \prod \limits_{k=1}^{N_{train}} \exp[\alpha_{Y_{train_{k}}}+\beta_{Y_{train_{k}}}Z_{t_ {k,Y_{train_{k}}}}]/N_{k}
\end{eqnarray}
Here, $\alpha$ is a vector corresponding to $(\alpha_1,\alpha_2,\alpha_3)$,  $\beta$ is a vector corresponding to $(\beta_1,\beta_2,\beta_3)$, and individual elements of a vector are indexed by subscript. So for instance, $Z_{t_ {k,Y_{train_{k}}}}$ refers to the $k_{th}$ element of the training latent vector of type $Y_{train_{k}}$. Moreover, the normalizing constant  $N_k$ is given by
\begin{eqnarray}
N_k &=& \sum_{l=1}^3 \exp[\alpha_{l}+\beta_{l}Z_{t_{k,l}}]
\end{eqnarray}
For notational brevity, denote the posterior distribution of the vector $Y_{pred}$,  $p^{*}(Y_{pred}|Y_{train}, X_{train}, X_{pred})$,  as $p^{*}(Y_{pred})$. Moreover, let $p$ be shorthand for $pred$ and $t$ be shorthand for $train$, and assume that $Z$ stands for the latent variables for the three compact object types. Then by the definition of conditional probability and marginalization, we have the following integral representation for $p^*(Y_{pred})$. 
\begin{eqnarray}
p ^*(Y_{pred}) &=& \int_{Z_{p},Z_{t},\alpha,\beta} p(Y_{p},Z_{p},Z_{t},\alpha,\beta|Y_{t},X_{t},X_{p}) dZ_pZ_td\alpha d\beta\\
                                                             &=& \int_{Z_{p},Z_{t},\alpha,\beta} p(Y_{p}|Z_{p},\alpha,\beta,-)p(Z_p|Z_{t},-)p(Z_{t},\alpha,\beta,|Y_{t},X_{t},X_{p})dZ_pdZ_td \alpha d\beta
\end{eqnarray}
Note that we purposely overload the $p(.)$ notation and use the dash symbol to represent ``all other variables", for less clutter. This decomposition suggests an iterative algorithm described in the appendix for sampling from the posterior distribution of $Y_{pred}$.

\section{Results}
Here, we present predictions of the compact object types of those systems in the test set discussed in Section 2 using the model and algorithm discussed in Section 3. The predicted class of an XRB is the one with the maximum estimated probability, an approach that can be justified from a decision theoretic view point because, in the discrete case, the posterior mode is a Bayes estimator under a 0-1 loss function. The class with the maximum estimated probability is mathematically equivalent to the class with the maximum number of posterior predictive draws for the class labels: i.e., the posterior predictive mode. 

Table 1 lists the predictions for the 6 systems whose classifications are known in addition to probability estimates and associated standard errors. Using this scheme, there are no misclassifications for this group of X-ray binary systems. Additionally, in Table 2 we include the predictions  and probability estimates for ``burster" non-pulsing systems, for which there are a number of wrong predictions: 4 out of 6 systems. In all cases these systems are mistaken for containing black holes.  Visual inspection of the data is consistent with this result because the regions these systems occupy interferes with the region defined by systems containing black holes: for instance, consider Figure 5 comparing a burster system that is misclassified as a black hole system to a burster system that is not misclassified, where there appears to be significantly more overlap with the black hole system training data for the misclassified system. It is also possible that some of these systems are misclassified as non-pulsing neutron star systems in the literature, yet significantly more scientific investigation must be performed in order to verify this possibility. In Table 3, we include predictions and probability estimates for unclassified or ambiguously classified XRB systems, for which notably GX 3+1 has a reasonably high estimated probability of being a non-pulsing neutron star system: .7674 with a standard error .0326.

It is important to note that the entire distribution of posterior predictive draws provides significantly more spatial information than a point estimate for compact system type. For instance, consider the systems Ser X-1 and Aql X-1, the first of which is a properly classified non-pulsing system, and the second of which is one that is improperly classified as a black hole. From Figure 6  there seems to be no question that Ser X-1 is indeed a non-pulsing neutron star system, since the proportion of posterior predictive draws is .9341 with a standard error of .0133. On the other hand, while Aql X-1 is not properly classified, as is evident in Figure 7, we do see some signal for non-pulsar. This is evidenced by the .3093 estimated probability of being a non-pulsing system, with standard error 0.0507. 

\section{Summary and Future Directions}
The main objective of this work has been to develop a probabilistic model for predicting the compact object type of CCI observations from an XRB and to use the predictions generated from this model to estimate the probabilities that the compact object type is a black hole, non-pulsing neutron star, or pulsar. We have shown that the model we have developed works reasonably well for this purpose based on the accurate classification of well known X-ray binaries, but we note that the model seems to make mistakes for the classification of bursters that are close to the boundary between black hole systems and non-pulsars in the CCI coordinate system. This suggests further investigation of these systems as well as the refinement of our approach, including the sampling of data, models, and algorithms used. It is also possible that some of these ``burster" systems are inappropriately classified, but more scientific investigation must be made before such a claim can be vigorously asserted.

In order to improve the predictive accuracy of our classification scheme, we can extend the imputation model by imposing a distribution on the Gaussian process parameters or else using a cross validation approach to fine tune these parameters. There is a growing literature on Gaussian process prediction that attempts to bypass the associated computational impediments, and so we would like to investigate these methods and their potential application to this problem further. An example of such an approach, for instance, is the INLA method introduced by \citet{rue09}. Additionally, the RXTE ASM data contains a large fraction of missing data due to the signal to noise threshold we employ, and so we may consider applying the framework of \citet{lit02} to model this missing data. The primary advantages of this approach are that it utilizes a Bayesian model for generating imputations,  which is consistent with our model for compact object prediction, and it may lead to more plausible predictions for the unknown compact object types. Additionally, we may want to consider different subsampling schemes besides the one we employ to ensure that they do not corrupt the inherent structure in the data set. Finally, in order to make the model more scientifically useful, we may want to include physically meaningful parameters; the inference of such parameters may explain the scientific reasons for the separation of observations into different regions by compact object type.

The CCI method uses measurements of X-ray intensity and color in two X-ray bands. This information will in general not only reflect on the properties of the source but also on the absorption of the intrinsic spectrum by the interstellar medium. ISM absorption will clearly affect the lowest energy band the most, and thus the soft color. However, at higher column densities, the hard color will be affected as well. We are developing a general method for correction of CCI plots given the sensitivity curve of an X-ray monitoring telescope and likely models of the spectral shape (Boroson et al., in preparation). The eROSITA telescope developed at the Max Planck Institute for Extraterrestrial Physics and due to be launched in 2016 has 20 times the sensitivity of the ROSAT/ASM in the low energy band and will be particularly beneficial to study the ISM. \citep{mer12}

We can extend our long-range study using data from past and present large field of view X-ray instruments such as MAXI \citep{mat09}, the HETE-WXM \citep{yos95} and BeppSAX-WFC \citep{boe97}.  Current and planned X-ray telescopes with high sensitivity such as Chandra \citep{wei02}, XMM \citep{mas95}, and eROSITA \citep{mer12} will enable us to apply our methodology to XRBs of much lower luminosity.

Finally, we reiterate that the \texttt{R} code we have written to make predictions for this analysis ought to be be applicable to other CCI data sets quite easily, and so we have provided it for public use.  

\acknowledgments

ICHASC is acknowledged for their helpful feedback. Additionally, GG and SDV would like to acknowledge partial support through a Smithsonian Institution CGPS grant to SDV.

{\it Facilities:} \facility{Harvard-Smithsonian Center for Astrophysics} \facility{Harvard University Odyssey Supercomputer}.

\appendix
\section{Algorithm description}
As described by \citet{gel13}, prediction in the Bayesian paradigm essentially follows the following iterative scheme: first draw from the posterior distribution of model parameters and latent variables through a Monte Carlo simulation, and then draw from the predictive distribution of interest, which in our case is that of $Y_{pred}$, conditional on these draws. Adapting this general strategy for our problem, Equations (4) and (5) from Section 3  suggest the following iterative algorithm for sampling from the posterior predictive distribution for the compact object type.
\begin{itemize}
\item Sample from the posterior distribution $p(Z_{train},\alpha,\beta,|X_{train},X_{pred},Y_{train})$ using elliptical slice sampling due to the joint multivariate normal distribution of $(\alpha, \beta,Z_{train})$. The method of elliptical slice sampling was introduced by \citet{mur10}.
\item Sample the posterior latent variables at the prediction points, $Z_{pred}$, using the conditional multivariate normal distribution of $p(Z_{pred}|Z_{train},-)$, which has mean $\Sigma_{pred,train}\Sigma_{train,train}^{-1}Z_{train}$ and covariance matrix $\Sigma_{pred,pred}-\Sigma_{pred,train}\Sigma_{train,train}^{-1}\Sigma_{train,pred}$ due to fundamental properties of conditional MVN distributions. $\Sigma$ is the covariance matrix for the latent variables, $pred$ corresponds to test prediction points, and $train$ corresponds to training points.
\item Sample from $Y_{pred}$ from a multinomial distribution conditional on the posterior latent draw of $Z_{pred}, \alpha,\beta$, where as aforementioned the probability for $Y_{pred_k}$ to be of type $l$ is proportional to $\exp[\alpha_l+\beta_lZ_{pred_{k,l}}]$. 
\end{itemize}
Additionally we set $\sigma^2 = 1$ and $\phi$ = 0.1. \footnote{More judicious ways of selecting these parameters are discussed in Section 5.}

Elliptical slice sampling is a Monte Carlo algorithm developed to simulate from a posterior probability distribution where the prior distribution is jointly multivariate normal, a condition that holds in our model as discussed in Section 3. As explained by \citet{mur10}, this is a scenario where traditional Monte Carlo methods applied within a Bayesian context, such as Gibbs sampling or Metropolis-Hastings, perform poorly. Routines to implement elliptical slice sampling and draw from the posterior distribution of $Z_{pred}$ and $Y_{pred}$ were written in the R programming language using the Rcpp, RcppEigen and RcppArmadillo packages for the efficient inline implementations of linear algebraic routines in C++ \citep{r15, bat13, edd13, edd14, sky15}.  Additional packages used in the testing and development of this code were mvtnorm and MASS \citep{genz14, ven02}. As discussed by \citet{mur10}, the computational impediments of elliptical slice sampling stem primarily from determining the Cholesky decomposition of and inverting a multivariate normal covariance matrix. RcppEigen and RcppArmadillo provide efficient implementations for determining the Cholesky decomposition and performing matrix inversion that can be conveniently included directly within R code. This code, along with the RXTE ASM data, is freely available at  \texttt{https://github.com/ggopalan/XRay-Binary-Classification}.

\clearpage

\begin{figure}
\plotone{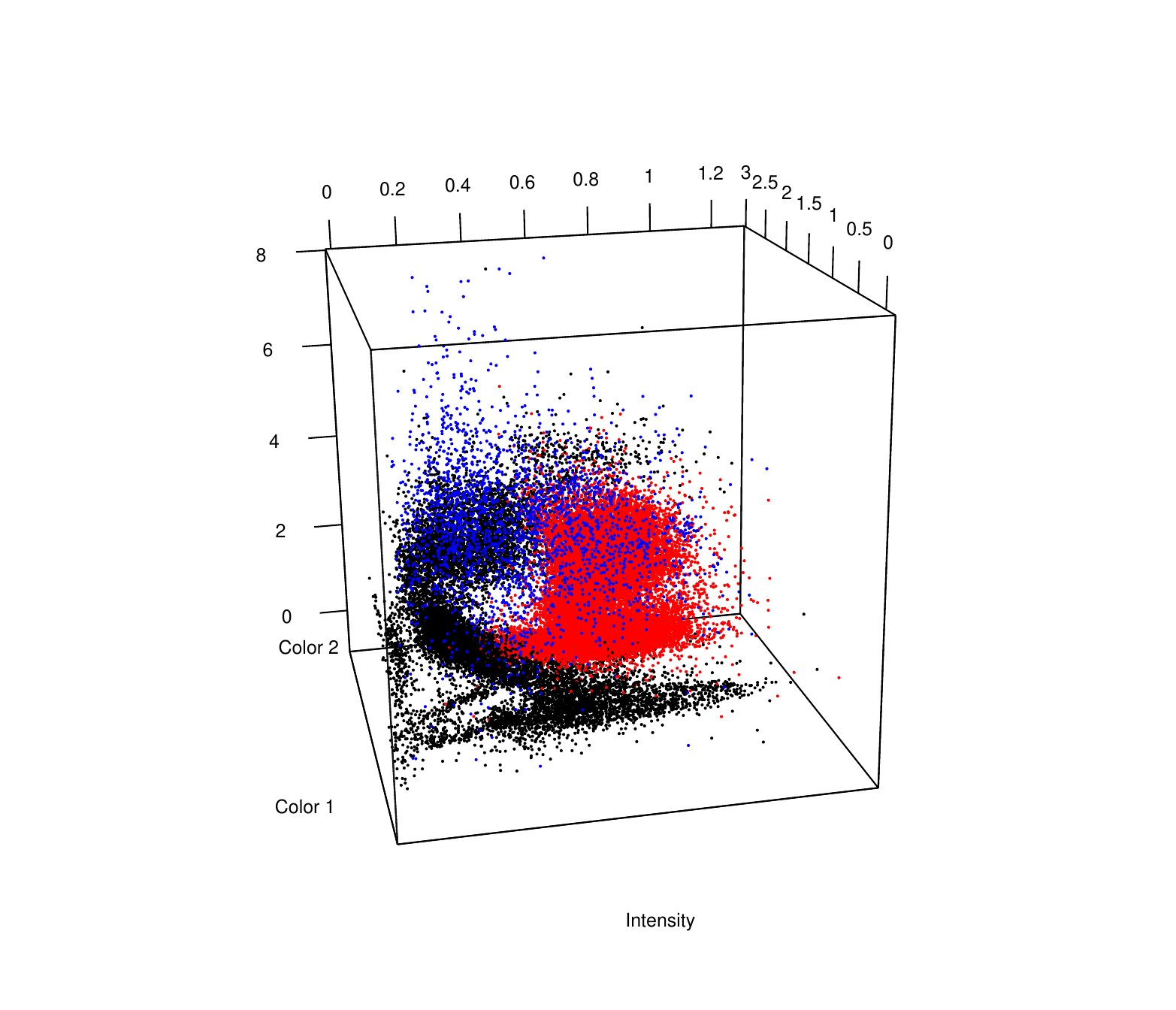}
\caption{Visualization of RXTE ASM data for 24 XRBs over 13 years including only the $5\sigma$ threshold values. Each individual point is the one day average of the CCI data from one of the 24 systems. Black points are observations from  black hole XRB systems, red points are observations from non-pulsing neutron star XRB systems, and blue points are observations from pulsing neutron star XRB systems. The general pattern is that observations from different system types separate geometrically in this CCI coordinate system. (Note that since CCI coordinates are defined in terms of ratios of counts, they are unitless.)}\label{fig1}
\end{figure}

\clearpage

\begin{figure}
\plotone{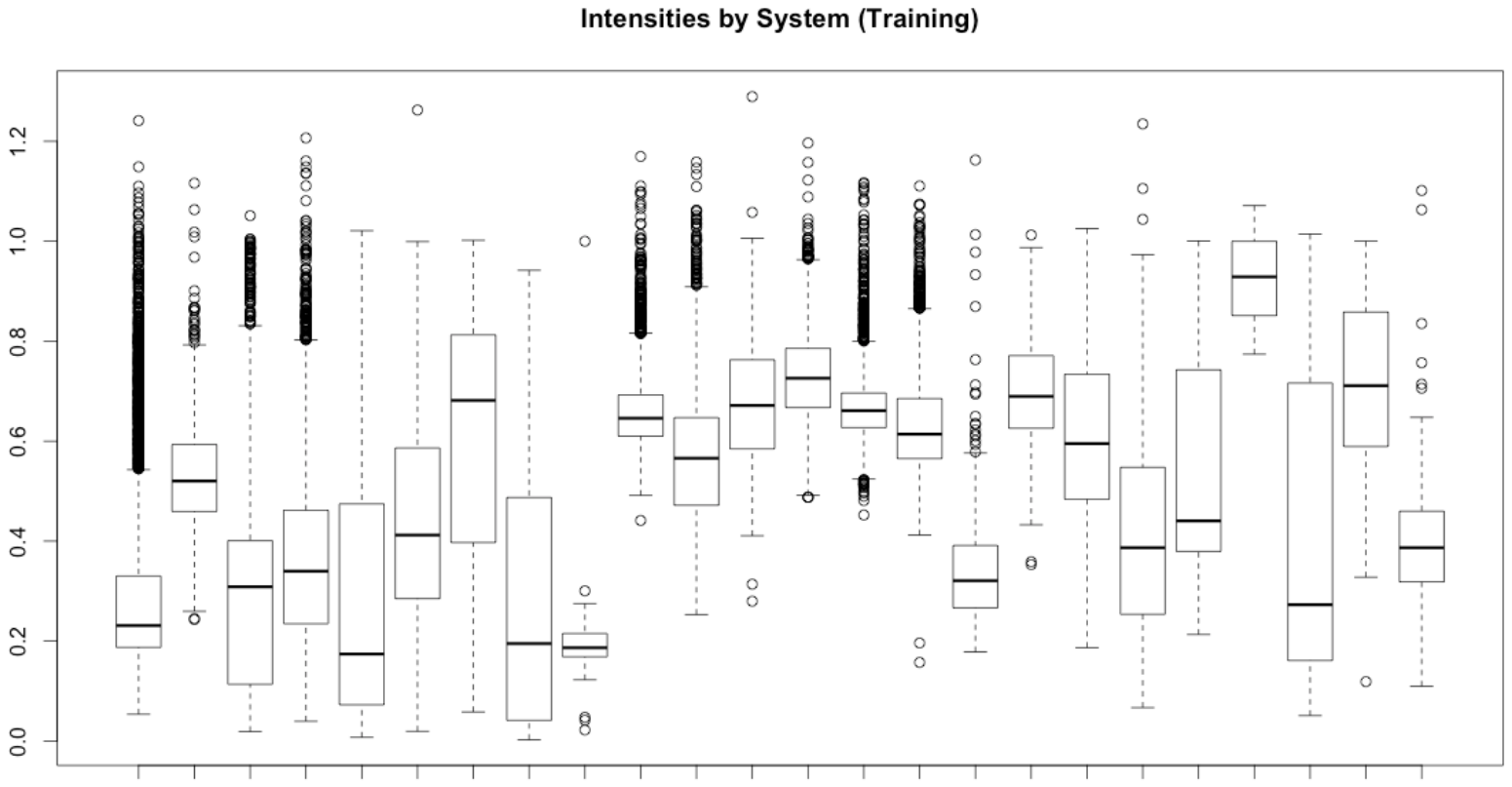}
\caption{Visualization of intensities for observations by each of the 24 systems within the training data set. Systems 1-9 are black hole systems, 10-15 are non-pulsing neutron star systems, and 16-24 are pulsing neutron star systems. The wide variability in intensities across systems may explain the wide variability for the number of observations of each system above the signal to noise threshold, since fainter measurements tend to be noisier.}\label{fig2}
\end{figure}

\begin{figure}
\plotone{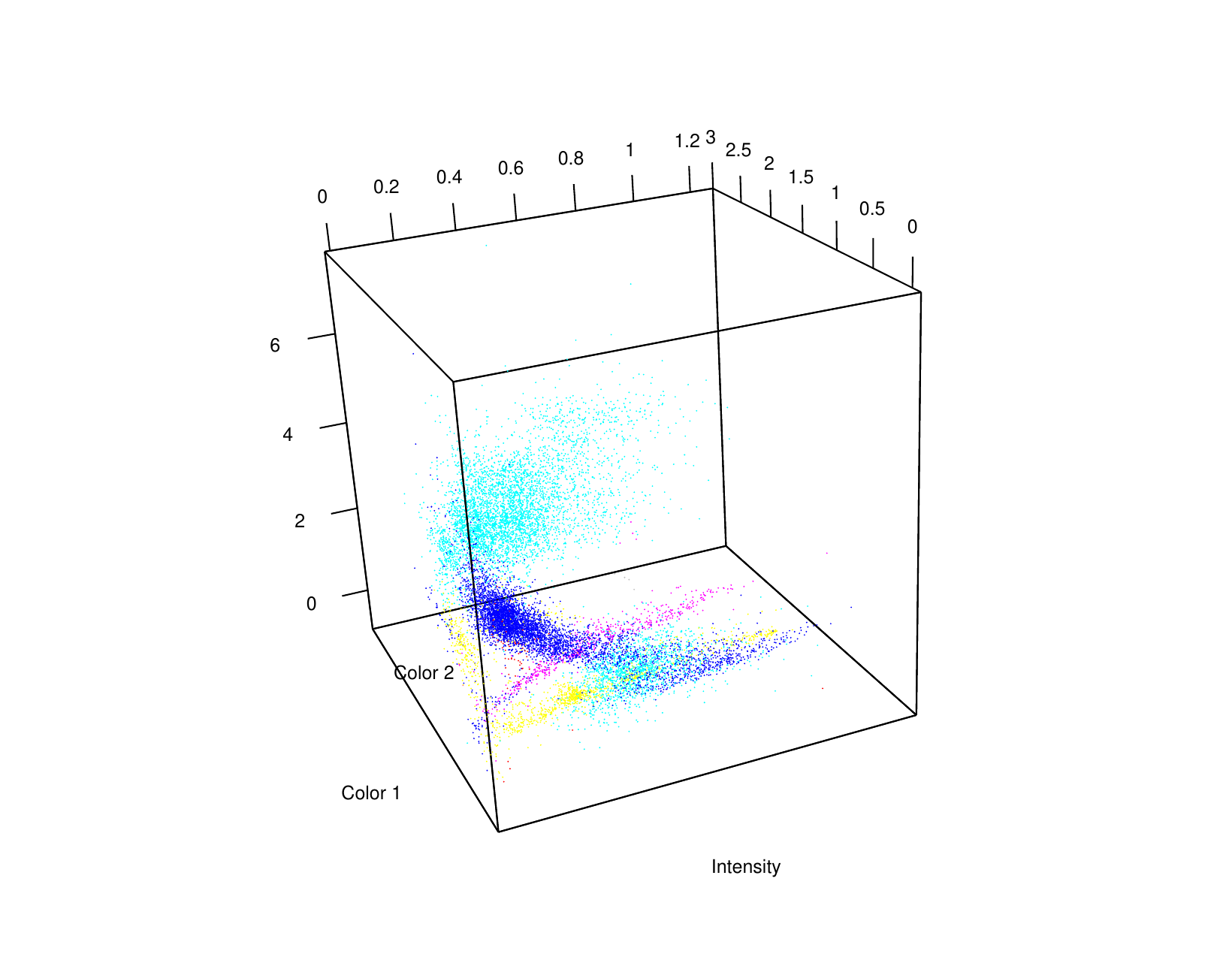}
\caption{An illustration of the wide variability in CCI data between different systems that contain a black hole, where each color represents data from a different system.}\label{fig3}
\end{figure}

\begin{figure}
\plottwo{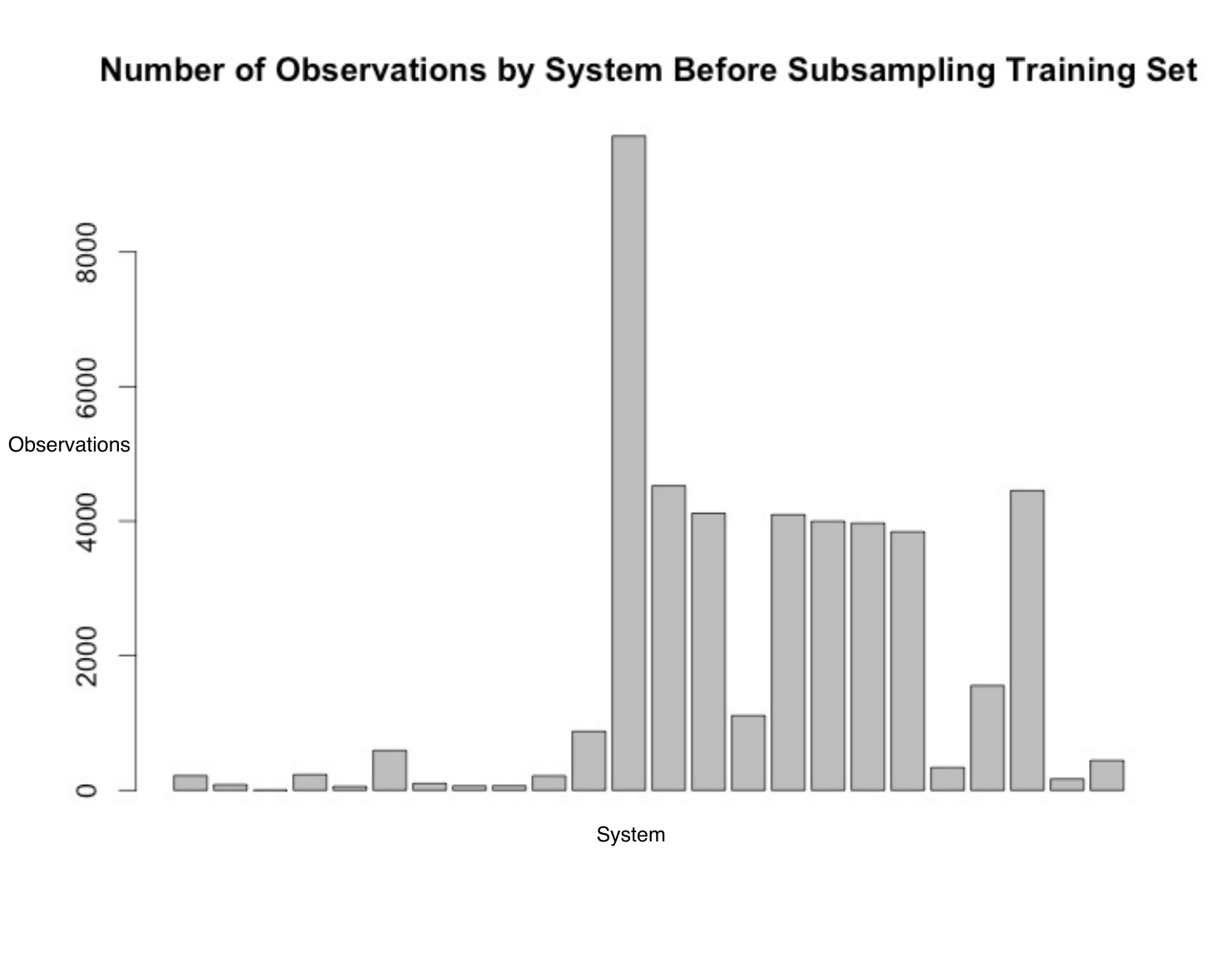}{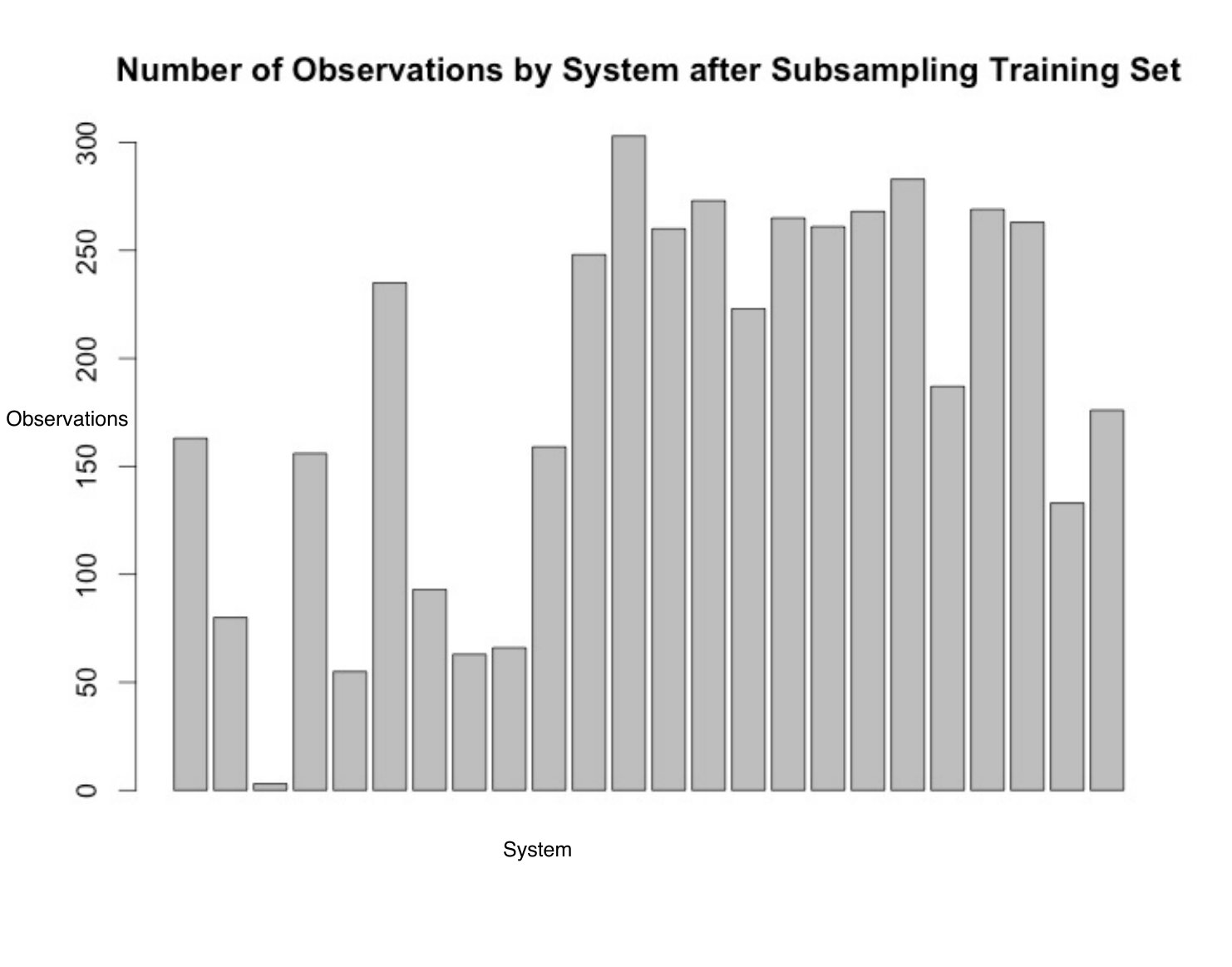}
\caption{A comparison of the distribution of the number of observations by system in the training set before and after subsampling.} 
\label{fig3}
\end{figure}

\begin{figure}
\plottwo{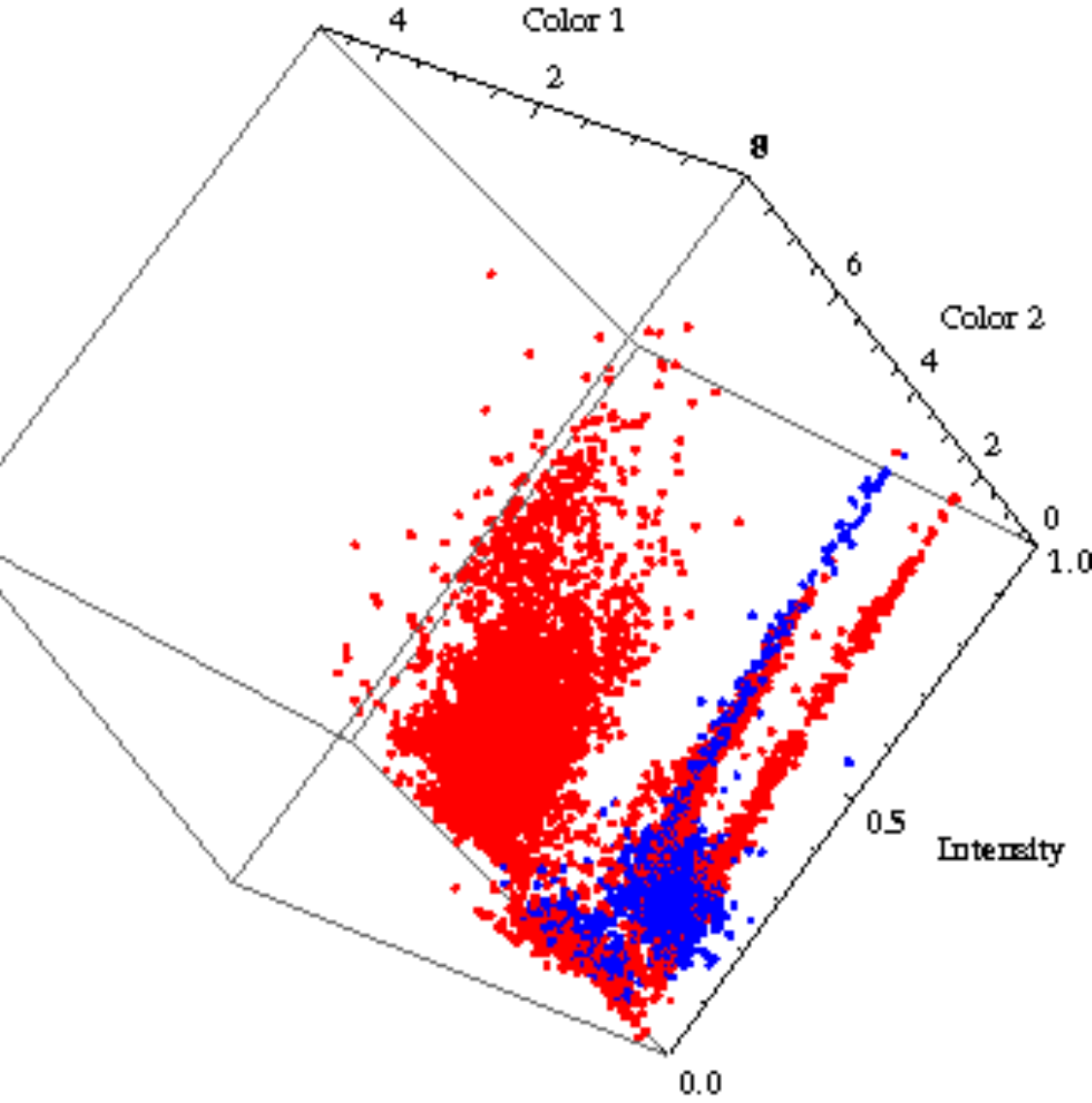}{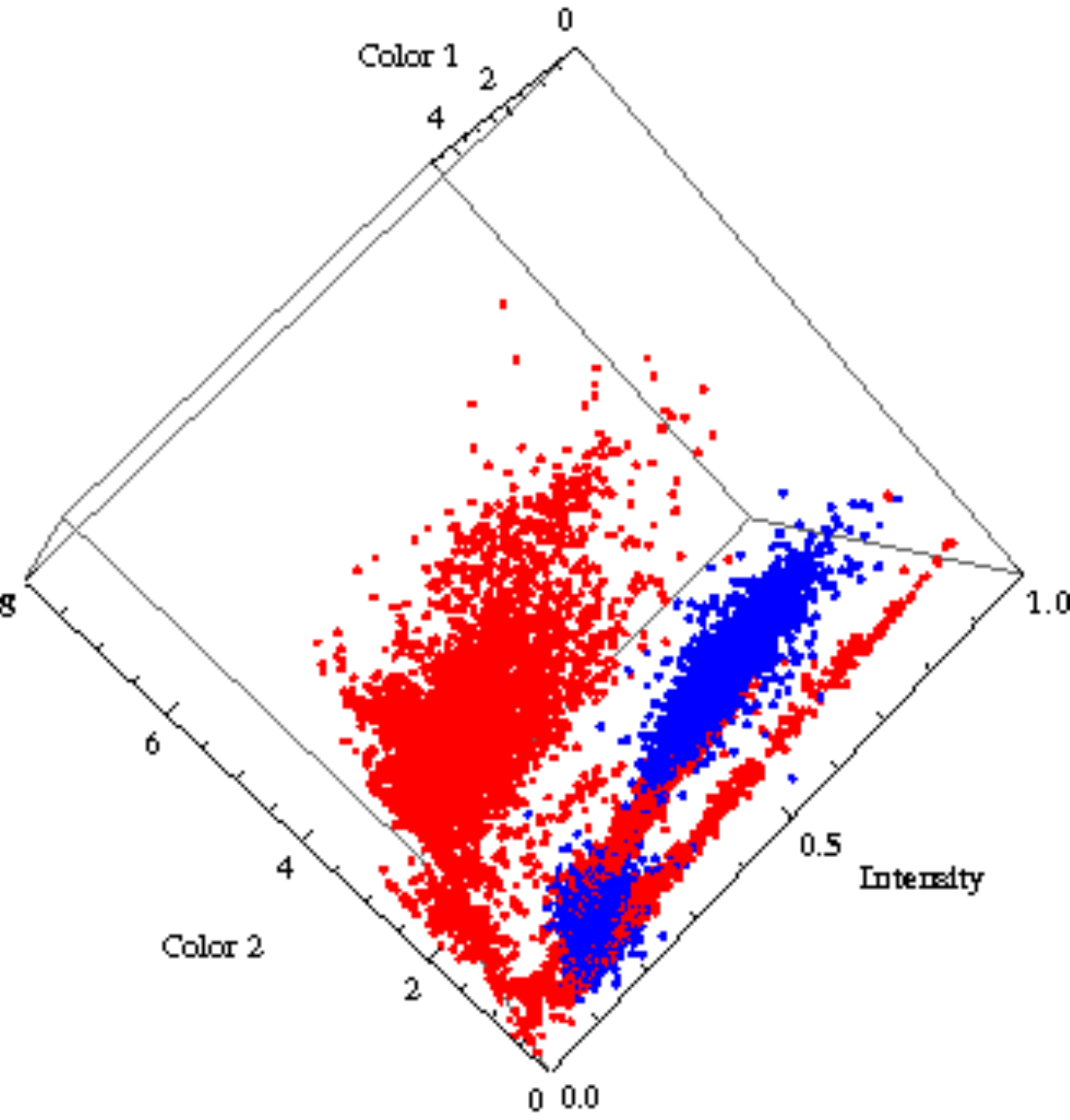}
\caption{Left: An example of a burster system in blue,  Aql X-1, improperly classified as a black hole system by the algorithm with comparison to black hole training data in red. Right: An example of a burster system in blue, Ser X-1, properly classified as a non-pulsing system by the algorithm with comparison to the black hole training data in red.}
\label{fig4}
\end{figure}

\begin{figure}
\plottwo{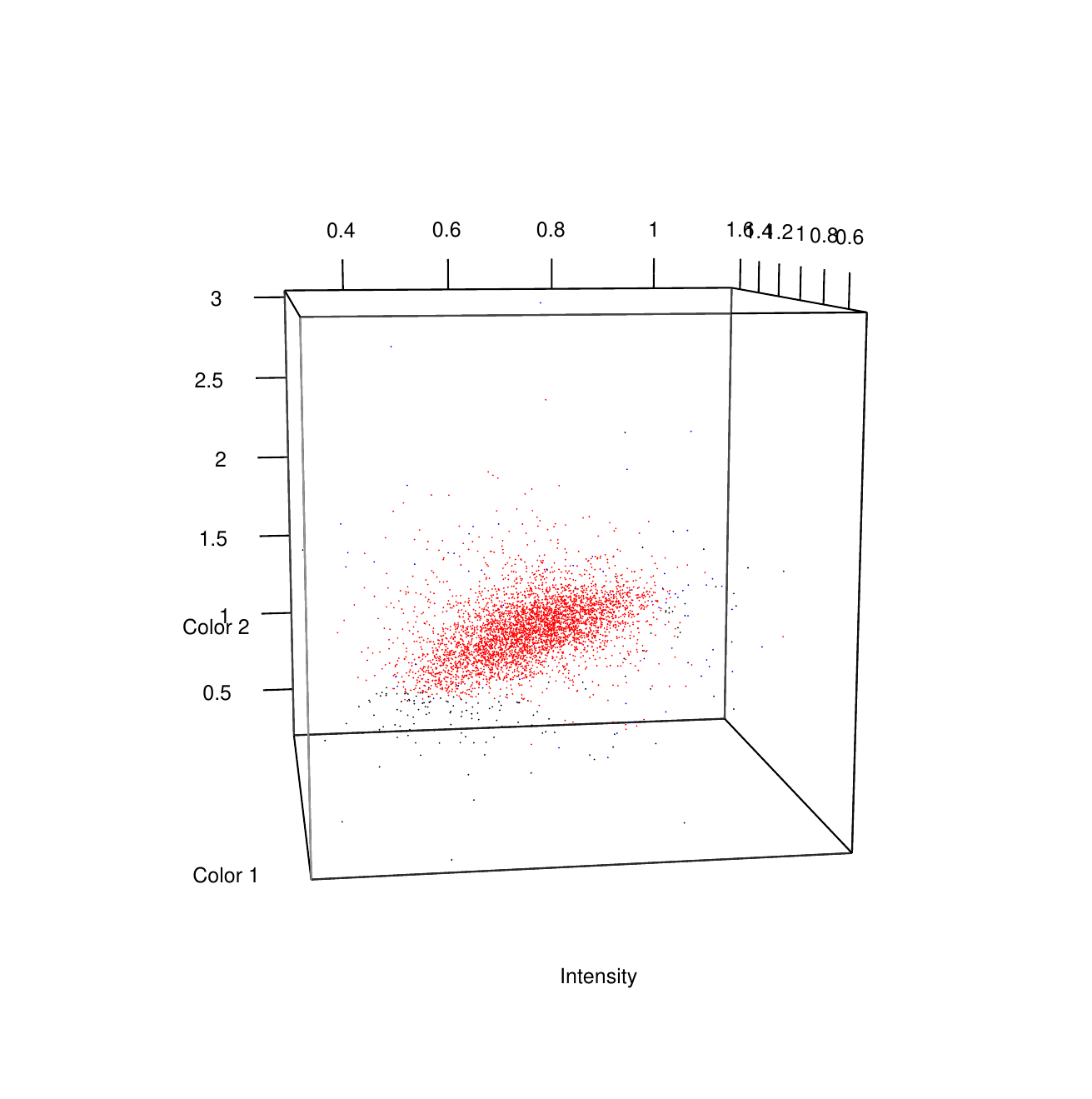}{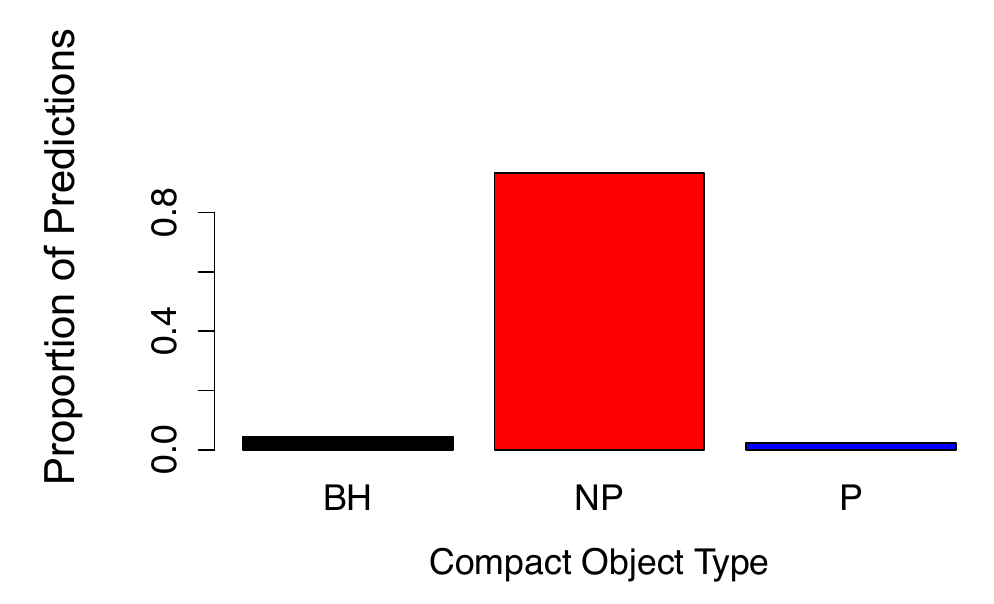}
\caption{An example of a non-pulsing neutron star system (Ser X-1) that is properly classified by the classifier. The left indicates all of the observations from Ser X-1 and the associated predictions of each observation, with the mode of posterior predictive draws taken to be the prediction for a given observation. The bar chart on the right illustrates the probabilities estimated that Ser X-1 is of each of the three classes.}
\end{figure}

\begin{figure}
\plottwo{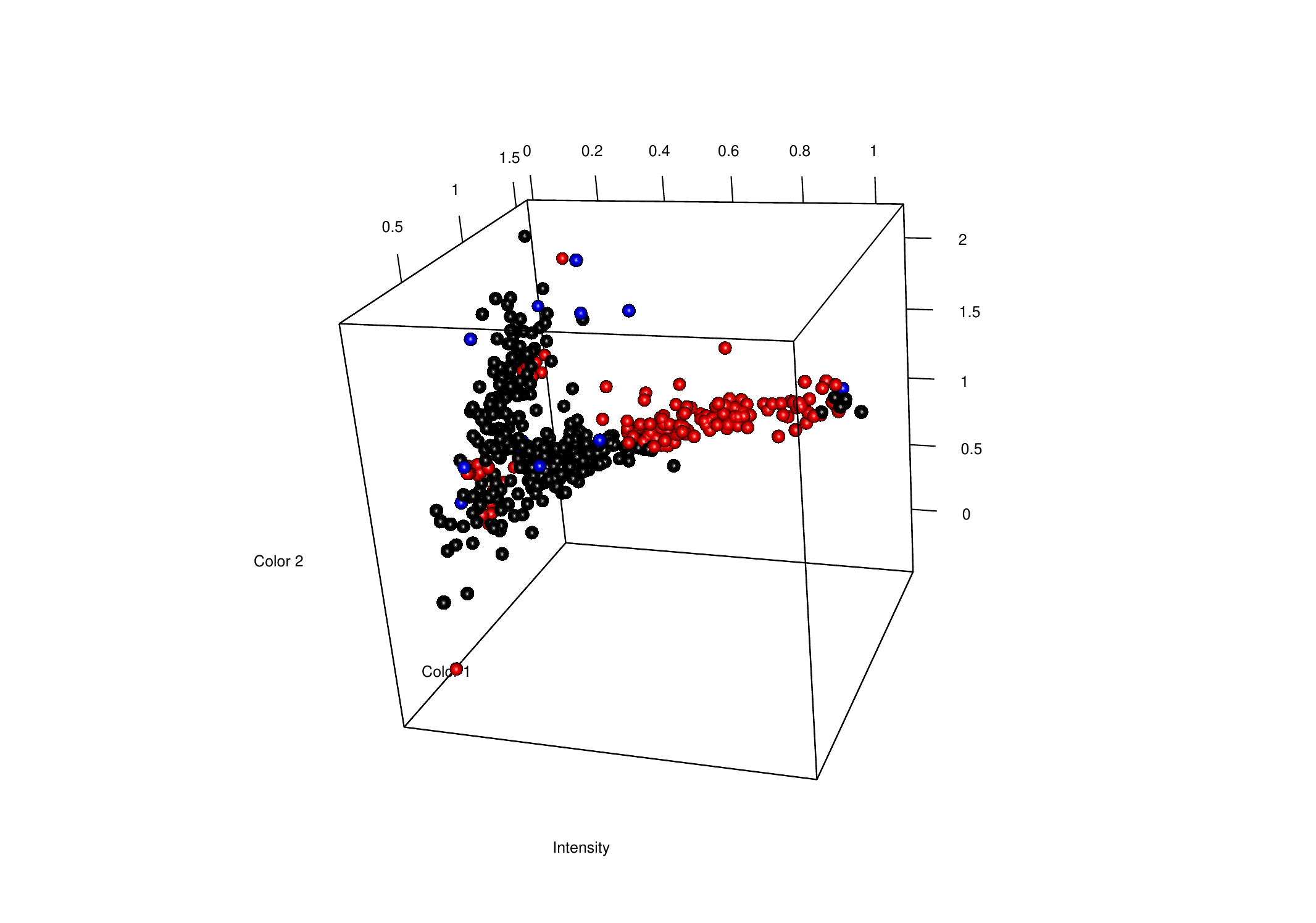}{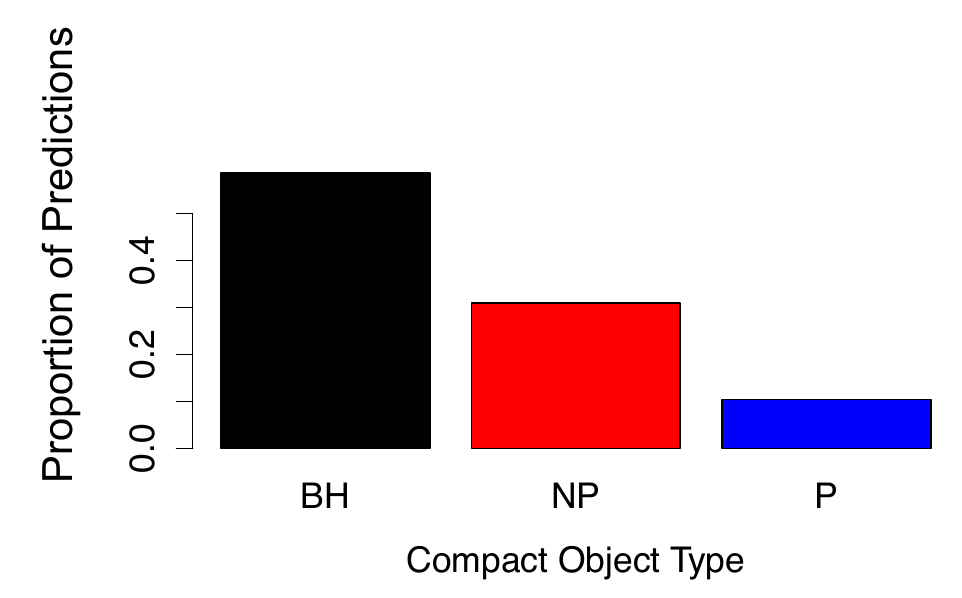}
\caption{An example of a non-pulsing neutron star system (Aql X-1) that is improperly classified by the classifier, mistaken for a black hole system. There appears to be some signal for a non-pulsar system, however.}
\end{figure}

%
%TABLES
%

\begin{deluxetable}{crrrrrrrrrrr}
\tabletypesize{\scriptsize}
\rotate
\tablecaption{Probability estimates and predictions for compact object type of previously classified XRBs}
\tablehead{
\colhead{System} & \colhead{$P(BH)$} & \colhead{SE} & \colhead{$P(Nonpulsar)$} & \colhead{SE} & \colhead{$P(Pulsar)$} &  \colhead{SE} & \colhead{Prediction }& \colhead{Actual Class} }
\startdata
 GX 5-1 & .0853 & .0519 &  .8285 & .0756 &  .0863 & .0259 &  NP & NP\\
 1744-28  & .2626 & .0978 & .1244 & .0570  & .6129 &.1072 & P & P \\
 0656-072  & .0782 & .0565 & .0659 & .0472 & .856 &  .0842 & P & P \\
 0535+262  & .2164 & .0728 & .1312 & .0562 & .6525 & .0883 &P & P \\
 0115+634  & .2179 & .0780 & .1425 & .0539 & .6396 & .1057 & P & P \\
 LMC X-3 & .8585 & .0648 & .0762 & .0347 & .0654 & .0379& BH & BH \\
\tableline
\enddata
\tablecomments{We demonstrate the estimated probabilities and predictions for 6 XRBs whose classification have been established in the literature, along with their standard errors. All of these systems are properly classified. The predicted class is the one with maximum estimated probability, which is equivalent to the mode of the posterior predictive draws; see Section 4 for more discussion of the justification of this choice.}
\end{deluxetable}

\begin{deluxetable}{crrrrrrrrrrr}
\tabletypesize{\scriptsize}
\rotate
\tablecaption{Probability estimates and predictions for compact object type of ``burster" XRBs}
\tablehead{
\colhead{System} & \colhead{$P(BH)$} & \colhead{SE} & \colhead{$P(Nonpulsar)$} & \colhead{SE} & \colhead{$P(Pulsar)$} &  \colhead{SE} & \colhead{Prediction }& \colhead{Actual Class} }
\startdata
 Ser X-1 & .0441 & .0081 & .9341 & .0133  & .0218 & .0077 &NP & NP\\
 Aql X-1 & .5869 &.0692 & .3093 & .0507 & .1038 & .0411 & BH & NP \\
 1916-053  & .7683 & .1328 & .1010 & .0810 & .1307 & .0765 & BH & NP \\
 1608-522  & .4919 & .0678 &.3535 & .0492 &.1545 &.0310 & BH & NP \\
 1254-69  & .2472 & .0231 & .6330 & .0246 & .1200 & .0181 & NP & NP \\
 0614+091 & .7062 & .0560 & .1720 & .0441 & .1217 & .0392 & BH & NP \\
\tableline
\enddata
\tablecomments{We demonstrate the estimated probabilities and predictions for 6 XRBs that are classified as ``bursters" along with their standard errors. Four of these six ``bursters" are improperly classified as black holes, and the rest are correctly classified. The predicted class is the one with maximum estimated probability, which is equivalent to the mode of the posterior predictive draws; see Section 4 for more discussion of the justification of this choice.}
\end{deluxetable}

\begin{deluxetable}{crrrrrrrrrrr}
\tabletypesize{\scriptsize}
\rotate
\tablecaption{Probability estimates and predictions for compact object type of unclassified or ambiguously classified XRBs}
\tablehead{
\colhead{System} & \colhead{$P(BH)$} & \colhead{SE} & \colhead{$P(Nonpulsar)$} & \colhead{SE} & \colhead{$P(Pulsar)$} &  \colhead{SE} & \colhead{Prediction }}
\startdata
 1900-245 & .4858 & .0989 & .2334 & .0736 & .2808 & .0846 & BH\\
 GX 3+1  & .0757 & .0284 & .7674 &.0326 & .1569 & .0210 & NP\\
 1701-462 & .5025 & .1107 & .2179 & .0642 & .2796 & .0698 &BH \\
 1700-37  & .1574 & .0650 & .1497 & .0503 & .6926 & .0749 & P \\
 1636-53  & .3619  & .0304 & .5411 & .0235 & .0969 & .0263 & NP \\
\tableline
\enddata
\tablecomments{We demonstrate the estimated probabilities and predictions for 5 XRBs whose classifications are unknown or ambiguous, along with their standard errors. The predicted class is the one with maximum estimated probability, which is equivalent to the mode of the posterior predictive draws; see Section 4 for more discussion of the justification of this choice.}
\end{deluxetable}

\end{document}